\documentclass[apj]{emulateapj}

\usepackage{graphicx,color}
\usepackage{amssymb,txfonts}
\usepackage{epsfig,natbib}
\usepackage{multirow}
\usepackage{rotating}
\usepackage{sidecap}
\usepackage{hyperref}

\renewcommand{\tabcolsep}{5pt}

\DeclareMathAlphabet{\mathitbf}{OML}{cmm}{b}{it}
\DeclareMathAlphabet{\mathf}{OML}{cmm}{c}{sl}

\newcommand{\eg}{e.g.,}

\slugcomment{Draft \today}
\shorttitle{Trigger mechanism of sub-flares in a braided magnetic structure}
\begin{document}

\title{Trigger Mechanism of Solar Subflares in a Braided Coronal Magnetic Structure}
\author{Sanjiv K. Tiwari\altaffilmark{1}, Caroline~E.~Alexander\altaffilmark{1}, Amy R. Winebarger\altaffilmark{1}, Ronald L. Moore\altaffilmark{1} } 
\email{sanjiv.k.tiwari@nasa.gov}
\altaffiltext{1}{NASA Marshall Space Flight Center, Mail Code ZP 13, Huntsville, AL 35812, USA}

\begin{abstract}
Fine-scale braiding of coronal magnetic loops by continuous footpoint motions may power coronal heating via nanoflares, which are spontaneous fine-scale bursts of internal reconnection. An initial nanoflare may trigger an avalanche of reconnection of the braids, making a microflare or larger subflare. In contrast to this internal triggering of subflares, we observe external triggering of subflares in a braided coronal magnetic field observed by the {\it High-resolution Coronal Imager (Hi-C)}.  
We track the development of these subflares using 12 s cadence images acquired by {\it SDO}/AIA in 1600, 193, 94 \AA, and registered magnetograms of {\it SDO}/HMI, over four hours centered on the {\it Hi-C} observing time. These data show numerous recurring small-scale brightenings in transition-region emission happening on polarity inversion lines where flux cancellation is occurring. We present in detail an example of an apparent burst of reconnection of two loops in the transition region under the braided coronal field, appropriate for releasing a short reconnected loop downward and a longer reconnected loop upward. The short loop presumably submerges into the photosphere, participating in observed flux cancellation. A subflare in the overlying braided magnetic field is apparently triggered by the disturbance of the braided field by the reconnection-released upward loop. At least 10 subflares observed in this braided structure appear to be triggered this way. How common this external trigger mechanism for coronal subflares is in other active regions, and how important it is  for coronal heating in general, remain to be seen.  

\end{abstract}

\keywords{Sun: corona --- Sun: flares --- Sun: magnetic fields --- Sun: photosphere --- Sun: transition region}

\section{INTRODUCTION}\label{intro}
The corona, the outer solar atmosphere, is somehow heated to a million Kelvin or more, hundreds of times hotter than the photosphere, the solar surface. Active-region (AR) coronae can be hotter than quiet-Sun (QS) and coronal-hole (CH) coronae by a factor of 4 to 10 \citep{zirk93}. Although several mechanisms have been proposed to account for coronal heating, e.g. wave heating, fine-scale magnetic flare heating, heating by chromospheric and coronal jets and spicules, etc, how the corona is heated remains undetermined \citep[see \eg][]{wals03,asch04,klim06,depo07,vanb11,mcin11,wede12,real14}. Some of the more recent observations support a wave heating mechanism for the QS and CH coronae \citep{depo07,mcin11,hahn12,hahn13}. However, heating by fine-scale magnetic flares may be a predominant mechanism for AR coronae \citep[see e.g. the observations of][]{cirt13,wine13,test13,bros14}.   

Nanoflare reconnection in coronal braided magnetic fields is one of the most popularly argued heating mechanisms to account for the observed active-region coronal brightness \citep{parker72,parker83a,parker83b,parker88,prie02,gudi05,klim06,rapp08,reep13}. Continuous random footpoint motion builds up magnetic energy in the coronal field by stressing the field via braiding. Some of the stress energy is released as heat from current dissipation via nanoflare bursts of magnetic reconnection, each releasing $\le 10^{27}$ erg of magnetic energy. The accumulation of energy in braided loops may also lead to larger internally-triggered reconnection events \eg\ microflares or subflares with free energy release of $\sim$ 10$^{27}$ - 10$^{28}$ erg \citep{sves76,parker88}. The events studied in this Letter are subflares with energies of this order, that is, of the order of the energy of a GOES A-class flare \cite[see e.g. Appendix C of][]{tiw09a}.

In this Letter, we examine the triggering of subflares observed in a braided coronal-loop structure. Instead of finding the abovementioned anticipated internal triggering, we present clear evidence of external triggering of many subflares observed in a coronal braided magnetic structure. The first direct observational evidence of braided loops in an AR corona was given by \cite{cirt13} using high quality data obtained with the {\it High-resolution Coronal Imager (Hi-C}: \cite{koba14}). Launched on a rocket on 2012 July 11, {\it Hi-C} provided images of an AR at a spatial resolution of about 150 km and temporal cadence of $\sim$5 seconds in a narrow wavelength range centered at 193 \AA. The braided coronal structure studied in this paper has been corroborated by non-linear force-free field (NLFFF) modeling by \cite{thal14} using photospheric vector magnetograms.

\begin{figure*}[htp]
      \centering
      \includegraphics[width=\textwidth]{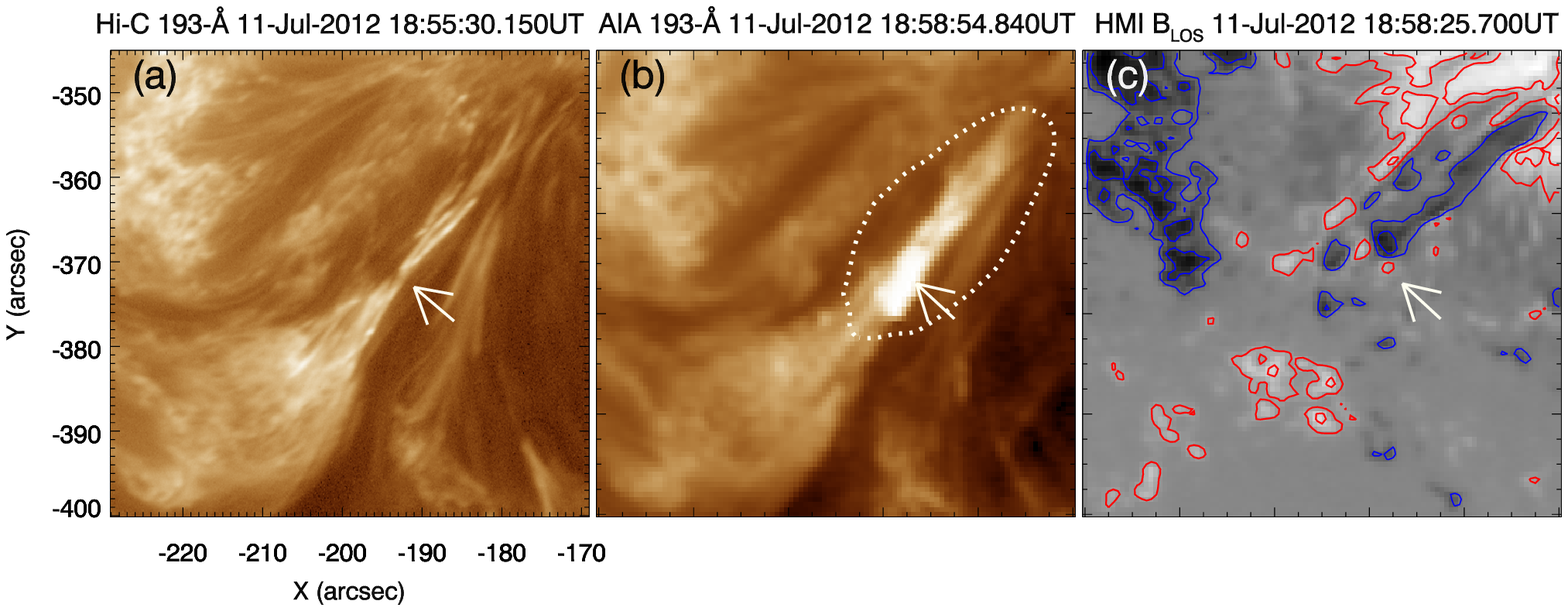}\vspace{0.1cm}
      \includegraphics[width=0.8\textwidth]{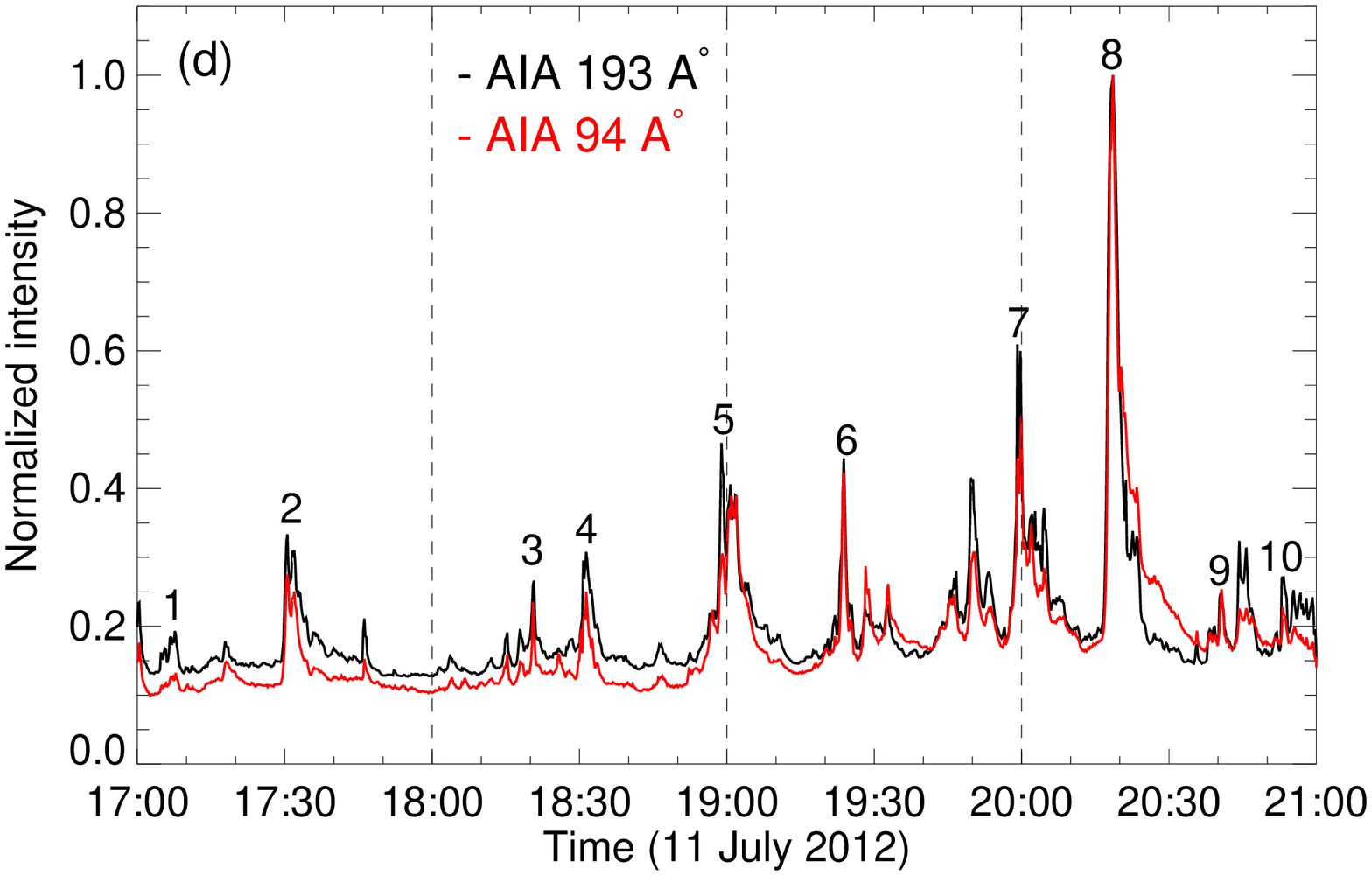}
      \caption{(a) An example Hi-C image with clear magnetic braided structure (indicated by an arrow) in which the 10 subflares, as numbered in (d) and listed in Table \ref{t1}, were observed during our observation interval of four hours. (b) An AIA image shown during a subflare in the braided structure immediately after the Hi-C span of observations. A dotted contour roughly outlines the braided coronal magnetic structure. (c) A LOS magnetogram with red/blue contours representing positive/negative polarity field of the level of $\pm$300, $\pm$700 G. (d) The AIA 193 and 94 \AA\ light curves for the braided area covered by the contour shown in (b). Each numbered peak is from the corresponding numbered subflare listed in Table \ref{t1}.}
      \label{hic+aia193+hmi}
\end{figure*}

We investigate the magnetic setting of our coronal subflares by using line-of-sight (LOS) magnetograms obtained by the Helioseismic Magnetic Imager (HMI: \citet{scho12}). \cite{falc97} and \cite{moor99} observed that enhanced coronal heating in ARs involves frequent microflaring, is concentrated in the magnetic field rooted close to polarity inversion lines (PILs) in the photosphere,  increases with increasing shear in the field at the PIL, and is modulated by some unknown additional factors. They suggested that a major factor was the rate of flux cancellation at the PIL, reconnection and submergence of short magnetic loops, driven by convective flows \citep{zwaa87} and/or by magnetic tension dominating over magnetic buoyancy \citep{rabi84}. However, further investigation of this suggestion has remained elusive in want of high spatial and temporal resolution cospatial photospheric and coronal data. Although flux cancellation on the photosphere has been studied separately \citep[see e.g.][for rates of flux cancellation in ARs]{park09}, this Letter presents new high-resolution observations showing in unprecedented detail the direct correspondence of coronal subflares to underlying small-scale flux cancellation.

\section{DATA SETS USED}\label{data}

We used four hours (17:00 - 21:00 UT) of movies from the Atmospheric Imaging Assembly \citep[AIA:][]{leme12}, in the 1600, 193, and 94 \AA\ channels, and LOS magnetograms from HMI, both onboard the {\it Solar Dynamics Observatory (SDO)} spacecraft. This time period covers two hours before and two hours after the $\sim$ 5 minutes of Hi-C 193 \AA\ observations. The pixel sizes of AIA and HMI are 0.6 and 0.5 arcsec, respectively. The temporal cadence of each AIA channel is 12 s and that of HMI is 45 s. 

The 1600 \AA\ passband on AIA is primarily lower-chromospheric continuum emission, but also covers the two C {\footnotesize IV}  lines near 1550 \AA\ formed at $T \approx$ 10$^5$ K in the lower transition region (TR). The short-term  brightenings in the 1600 \AA\ band have been found to be due to these C {\footnotesize IV} lines, and hence are from the lower TR \citep{leme12}. Thus, the brightening events seen in this band occur low in the magnetic field below the coronal braided structure and close above the field's evolving feet observed in the HMI magnetograms. 

The AIA 193 \AA\ and 94 \AA\ bands are both predominantly coronal \citep{leme12}. The 193 \AA\ band mostly detects Fe {\footnotesize XII} at about 1.5 MK, but has some response also to  TR emission from $2-3 \times10^5$ K plasma \citep{delz13,wine13}. The 94 \AA\ channel is centered on an Fe {\footnotesize XVIII} line ($6-8\times10^6$ K), but also detects some line emission from Fe ions formed at $\sim 1\times 10^6$ K \citep{warr12,delz13}; see also \cite{test12}. There is no known TR contamination in the 94 \AA\ channel.  
In this Letter, we use AIA 193 and 94 \AA\ movies to show the correspondence of brightening events (coronal subflares) in the braided coronal structure with nearly simultaneous events in the photosphere and lower TR. 

In Figure \ref{hic+aia193+hmi}(a), we display a Hi-C 193 \AA\ image from \cite{cirt13} that shows the braided coronal structure. We examined the evolution and events in the photosphere, TR, and corona in this field of view for $\pm$ 2 hours from the Hi-C flight time. We co-registered the images and magnetograms in Figure \ref{hic+aia193+hmi} and in the movie by using SolarSoft routines.

\section{TRIGGER MECHANISM OF SUBFLARES IN BRAIDED CORONAL MAGNETIC STRUCTURE}\label{trigger}
The braided coronal loop (Figure \ref{hic+aia193+hmi}a) displayed seventeen subflares, identified visually by examining nearly simultaneous intensity enhancements in 193 and 94 \AA, in the four hours of our observation period. The triggering of ten of these coronal subflares was clearer than in the other seven. The seven others overlapped these ten either in space and/or time, or were not so clearly triggered in the way that the ten were. In Figure 1(b), an AIA 193 \AA\ image displays a typical ongoing subflare event (indicated by an arrow). This is the subflare that we will present in detail. A LOS magnetogram near the time of the subflare is shown in Figure \ref{hic+aia193+hmi}(c). The magnetic field's complexity, displayed by the multiple bipolar structures and PILs under the coronal braided structure, is noteworthy.  

\begin{figure*}[htp]
      \centering
      \includegraphics[angle=90,width=\textwidth]{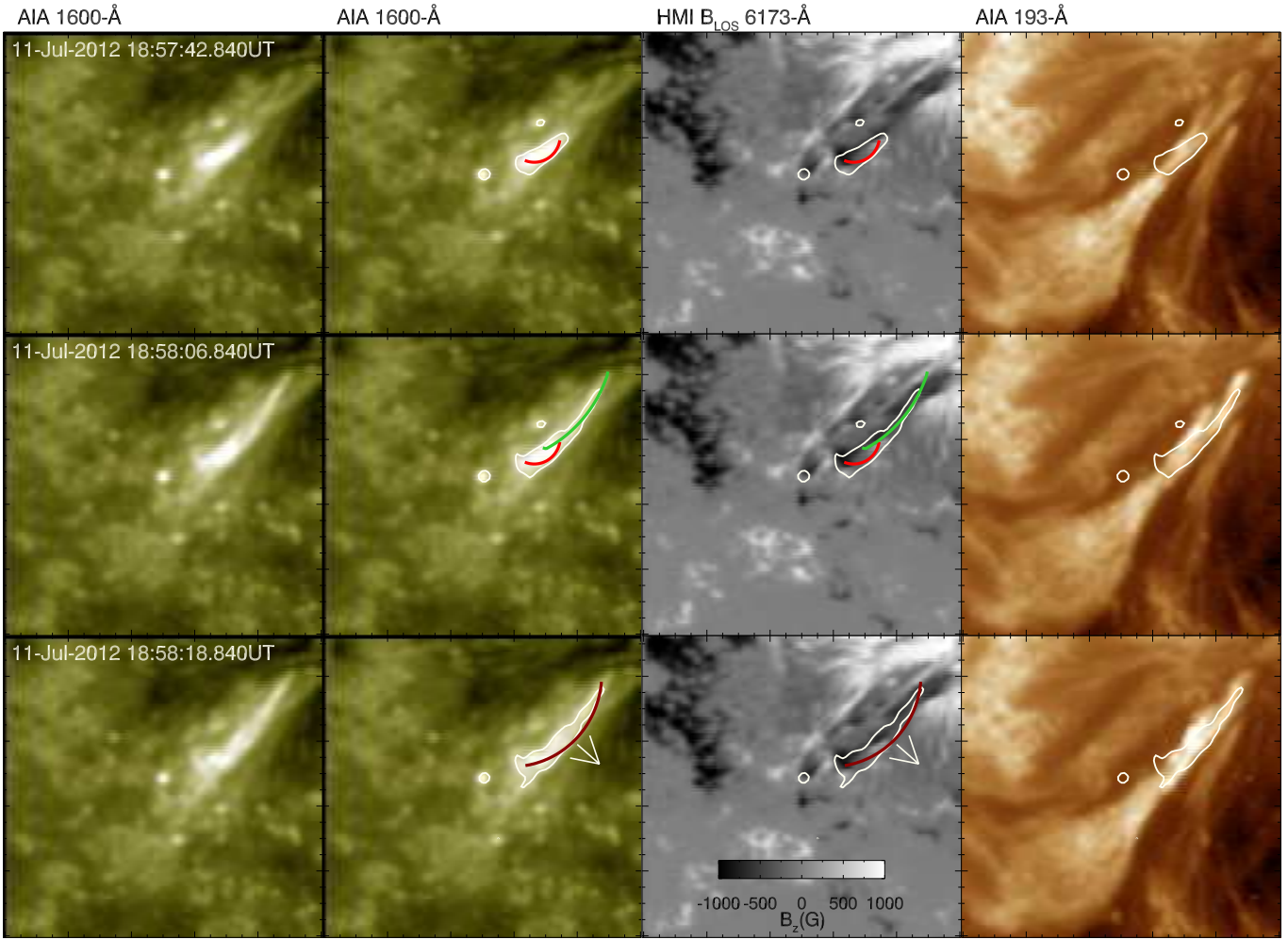}
 \caption{{\footnotesize Triggering of a typical coronal subflare in the braided structure, displayed in frames selected from the movie `a\_movie.mpeg'. The subflare started at 18:58:18 UT (see bottom right panel) and lasted about 48 s. Figure \ref{hic+aia193+hmi}(b) is a later frame during this subflare. Images in each row are at same time. The two columns on the left are images of AIA 1600 \AA. The two columns on the right are HMI B$_{LOS}$, and AIA 193 \AA\ images, respectively. Contours in each row are of the intensity enhancements in 1600 \AA, which are not shown in 1st column to keep the 1600 \AA\ images clear. The red, green and maroon colored arcs are to guide the eye and represent short, longer (not as bright) and resultant still-longer upward loop after magnetic reconnection, respectively. Arrows in the bottom row indicate the reconnected loop's inferred upward motion that triggers the subflare that is starting in AIA 193 \AA\ bottom right image. A cartoon presentation of this triggering event is shown in Figure \ref{cartoon}.}}
      \label{img_selected}
\end{figure*}

In the movie `a\_movie.mov' linked to Figure \ref{img_selected}, four panels, three of AIA channels: 1600, 193 and 94 \AA, and an HMI image of nearly same time as the AIA image, are displayed in each frame. The contours in the movie are the same for all panels in each frame, and outline the brightenings in the 1600 \AA\ channel. Note that most of the TR brightenings stand on PILs observed in the HMI images. We identified each coronal subflare by its brightening in the 193 and 94 \AA\ coronal movies. We found ten coronal subflares that were apparently triggered externally in the way described later in this section. These ten subflares are numbered in the light curve plots in Figure \ref{hic+aia193+hmi}(d), and the time and duration (the duration is the time difference between the first and last frames in which the subflare is visible) of each subflare are given in Table \ref{t1}. Each subflare's brightening in the AIA 193 \AA\ movie is labeled with that subflare's number, in a frame at the subflare's peak brightness. 

\begin{table} [h]
      \caption{Times of ten subflares seen in AIA 193 and 94 \AA\ that appear to be externally triggered in a similar way as our example subflare.}
     \vspace{-10pt}
      \begin{center}
            \begin{tabular}{cccc}
                  \tableline
                  Number &  Start~ time (UT) & End ~time (UT) & Duration (min) \\
                  \tableline
                  1 & 17:04:54 & 17:06:06 &  1.2\\
                  2 & 17:29:54 & 17:30:54 &  1.0\\
                  3 & 18:20:06 & 18:20:42 &  0.6\\
                  4 & 18:29:18 & 18:32:42 &  3.4\\
                  5 & 18:58:18 & 18:59:06 &  0.8\\
                  6 & 19:22:30 & 19:24:06 &  1.6\\
                  7 & 19:58:06 & 19:59:54 &  1.8\\
                  8 & 20:16:42 & 20:18:54 &  2.2\\
                  9 & 20:37:06 & 20:38:54 &  1.8\\
                10 & 20:51:54 & 20:53:18 &  1.4\\           
                  \tableline
            \end{tabular}
      \end{center}
      {\scriptsize Each subflare's number (as labeled in the AIA 193 \AA\ movie frame at the peak of each subflare), start time, end time, and duration, all obtained visually from the movie, are listed. 
}
      \label{t1}
\end{table}

We now focus on one of the ten subflares in Table \ref{t1}, subflare number 5. This subflare, shown in the Figure \ref{hic+aia193+hmi}(b), started at 18:58:18 UT and ended at 18:59:06 UT. In Figure \ref{img_selected}, we display a time sequence of images (selected from the movie) from two selected AIA wavelength bands, 1600 \AA\ and 193 \AA, and HMI magnetograms, for 36 s, from 18:57:42 UT to when the subflare was triggered. As in the movie, the contours on all three images of each row of the Figure \ref{img_selected} outline the intensity enhancements in 1600 \AA. In the movie, that most of the small-scale brightenings seen in 1600 \AA, and corresponding coronal brightenings in 193 \AA\ and 94 \AA, approximately center on PILs of the mixed-polarity flux in the photosphere, indicates that, not only on larger scales, as observed by \cite{falc97} and \cite{moor99}, but also on smaller scales, such as here, coronal heating events often happen in connection with PILs. 

Just before the example subflare started in the corona seen in 193 \AA\, a small loop gets noticeably brighter than the surroundings in 1600 \AA\ at 18:57:42 UT. Small-scale activity at this site is seen for about two minutes before as the loop-like structure brightens. Within 12 s, at 18:57:54 UT, an adjacent longer loop brightens in 1600 \AA. Within half a minute later, the 193 \AA\ movie and the 94 \AA\ movie both show the coronal subflare starting at 18:58:18 UT. The AIA 94 \AA\ images in the movie, which show only coronal plasma, are not shown in Figure \ref{img_selected}. The footpoint locations of the two 1600 \AA\ loops on the LOS magnetograms confirm that the above two bright ``loops" are magnetic loops, because opposite feet are rooted in opposite-polarity flux regions. Reconnection of these two loops, presumably causing the onset of the subflare brightening event in the 1600 \AA, plausibly made two reconnected loops, one shorter and downward, and the other longer and upward. The shorter loop could submerge owing to its strong curvature forces, consistent with the flux cancellation seen at that site. The larger loop, released upward, could interact with the overlying braided coronal magnetic structure and thereby trigger the coronal subflare that started there at 18:58:18 UT.

\begin{figure*}[htp]
      \centering
      \includegraphics[width=0.45\textwidth]{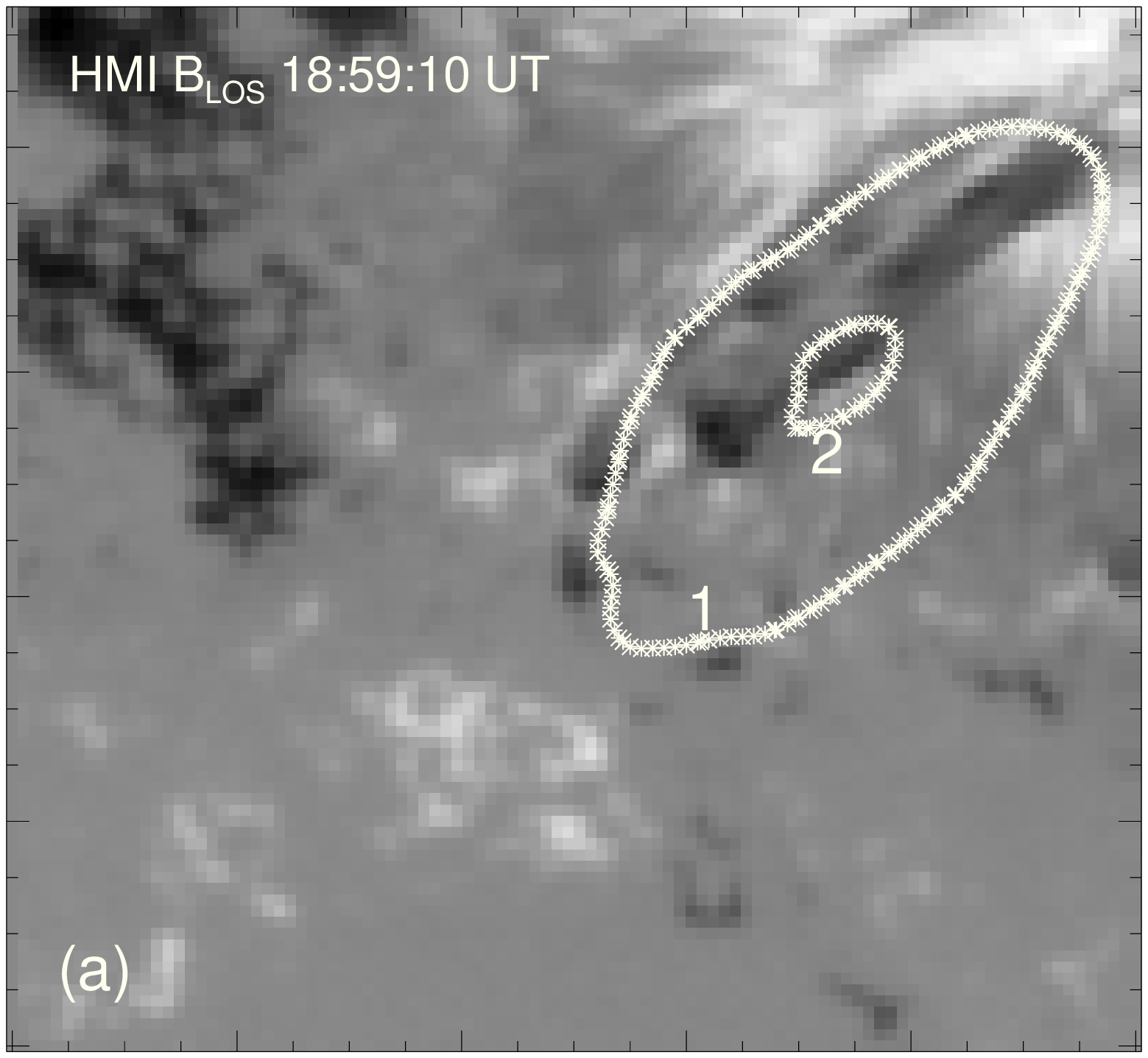} \hspace{0.3cm} \vspace{0.3cm}
      \includegraphics[width=0.52\textwidth]{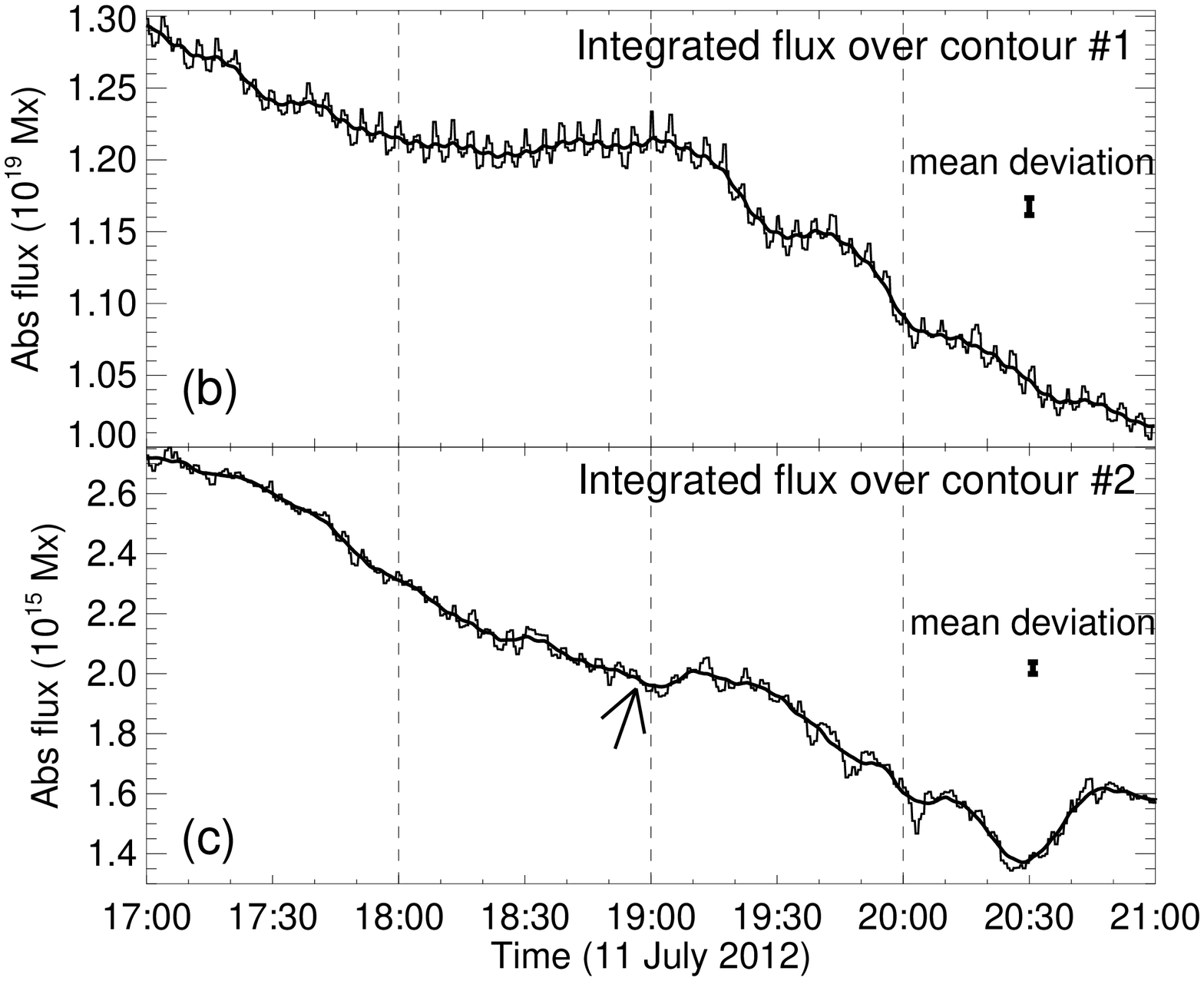} 
      \caption{\footnotesize (a). A LOS magnetogram with two contours named `1' and `2'. The contour `1' is the same as the dotted contour in Figure \ref{hic+aia193+hmi}(b) which roughly outlines the braided coronal region observed by Hi-C and is from the smoothed intensity of the 1600 \AA\ image at 18:58:18 UT.  The contour `2' outlines the region where the downward shorter reconnected loop should be submerging in the example subflare. (b). \& (c). Plots of the absolute magnetic flux integrated inside these two contours for four hours: the solid trend lines are from 10 minute smoothing; the fluctuations about them are real; their mean deviation is shown by the vertical bar. The arrow in (c) indicates flux cancellation at the time of the example subflare. The three vertical dashed lines are plotted, for reference, at 18:00, 19:00 and 20:00 UT.}
      \label{mag_flux}
\end{figure*}

We estimated the available free magnetic energy for this subflare to be $V B_\phi^2/8\pi$, where $V$ is the volume of the part of the coronal braided structure that brightened in the subflare, and $B_\phi$ is the azimuthal/twist component of the field in the flaring flux rope. We estimate from the subflare images and from the NLFF coronal field computed by \cite{thal14} from a vector magnetogram of the Hi-C AR that $V \sim 10^{27}$ cm$^{3}$ and $B_\phi \sim $100 G, which give $\sim 10^{29}$ erg. This amount of free energy is ample to produce subflares \citep{sves76}.  For the thermal energy, $3n_ek_BTV$, of the subflare, using plasma electron density of $\sim 10^{10}$ cm$^{-3}$ \citep[\eg][]{cirt13} and peak temperature (of 94 \AA) of 1$-$6 MK \citep[\eg][]{warr12,test12,cirt13}, we obtain $\sim 10^{28}$ erg, which is the nominal energy of a large microflare or small subflare \citep{sves76,parker88}.    

To demonstrate the magnetic flux cancellation taking place under the braided coronal structure, we show, in Figure \ref{mag_flux}(b) \& (c), evolution of the absolute flux inside the two contours `1' and `2' (labeled in Figure \ref{mag_flux}a.), respectively. The bigger contour outlines the whole area of the braided structure, and the smaller one outlines the site of the submergence of the short, downward reconnected loop. A general trend of decrease in the flux, at a typical rate of flux cancellation in ARs \citep[$\sim$ 10$^{18}$ Mx h$^{-1}$; see e.g.,][for measurements by MDI]{park09}, is obvious for four hours for the bigger contour, suggesting that the enhanced brightenings in the overlying TR and corona are caused by the flux cancellation in the photosphere. However, a nearly constant flux was seen within the larger contour at about the time of our example subflare event. To confirm the flux cancellation in the area of expected loop submergence, in the lower right panel we plot the absolute flux integrated over the area enclosed by contour `2'. The subflare occurred (arrow) near the end of a prolonged downward trend (solid curve) in the flux at the site of the TR brightening in the subflare onset.
This fits our idea of submergence of the small, downward reconnected loop.


The Figure \ref{mag_flux}(c) shows that the absolute flux within contour `2' decreased throughout our 4-hour observing span, except during two 15-20 minute intervals, starting at about 19:00 UT and at about 20:30 UT. In the 1600 \AA\ movie many sporadic brightenings occur at this site of flux cancellation, and often no coronal subflare is triggered. But each brightening at this site that was part of the triggering event for a coronal subflare occurred during the decreasing flux within contour `2'. This occurred for subflares 5, 7, and 8.     

Ten similarly triggered coronal subflares, bridging one or more PILs in the photosphere, were identified in the four hours of the movie (Table \ref{t1}). The triggering was similar in that the TR brightened at an underlying flux-cancellation site before (1-4 AIA frames) the coronal subflare started. The observations thus imply that each of these subflares was triggered externally by underlying reconnection accompanying flux cancellation as in the example presented here. 

\section{Discussion and Conclusions}
\begin{SCfigure*}
      \centering
      \includegraphics[trim=8.2cm 0cm 0.1cm 7.82cm,clip=true, width=0.6\textwidth]{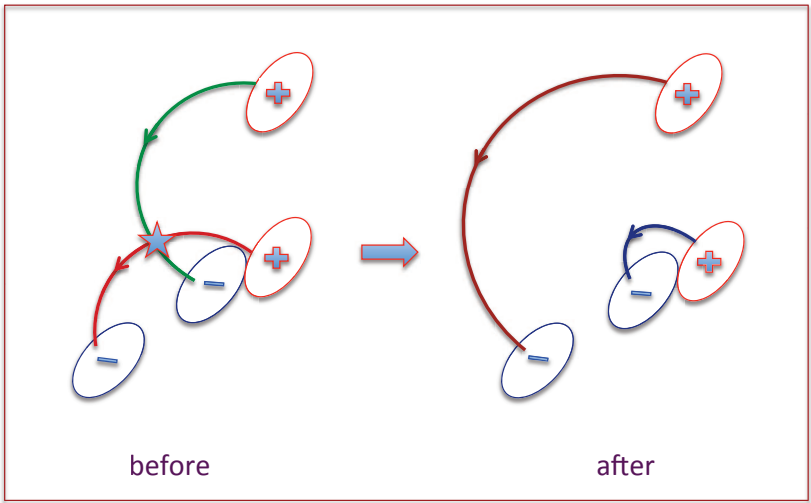}
      \caption{\protect\ A schematic sketch of the trigger of the subflare in the typical event, depicting the reconfiguration of the reconnecting field underlying the coronal braided magnetic structure. The TR brightening at the reconnection site is indicated by a star. A lower shorter (blue) and an upper longer (maroon) loop are formed by the reconnection.}
      \label{cartoon}
\end{SCfigure*}

We observed an external trigger mechanism that initiated at least ten out of seventeen subflares taking place in a braided magnetic structure in the Hi-C AR corona during the four hours of our observation. We presented in detail one coronal subflare that the observations show was triggered externally, and imply that the triggering was by an external reconnection event. In the presented example, this reconnection of the field underlying the braided coronal field would result in a shorter downward and a longer upward loop. The shorter loop would submerge, whereas the upward loop  would erupt and strike the overlying braided magnetic structure. This interaction plausibly triggered the coronal subflare, with an estimated energy of $\sim 10^{28}$ erg, in the coronal braided magnetic structure. 

Figures \ref{img_selected} and \ref{mag_flux}(c) together demonstrate crucial pieces of evidence of the inferred external trigger mechanism. In Figure \ref{cartoon}, we draw a sketch of the triggering external magnetic-reconnection event inferred from the observations of our example event, see all panels of Figure \ref{img_selected}. The red and green loops in the cartoon depict the two low-lying magnetic loops in the beginning of the reconnection. A star marks the location of reconnection in the lower TR. The two resultant loops of the reconnection are shown in the right part of the cartoon, where the small blue loop submerges into the photosphere and the flux cancellation at this site is seen as indicated in Figure \ref{mag_flux}(c). The larger maroon reconnected loop springs upward and interacts with the overlying braided structure, thereby triggering the observed subflare in the corona.


As described in the Introduction, it is commonly thought that spontaneous internal nanoflare reconnection in braided coronal fields triggers larger internal reconnection events (avalanches of nanoflares) seen as microflares or subflares \citep{sves76,parker88}. Our observations demonstrate that subflares in braided coronal fields can also be triggered externally, as in the example presented.  

The presence of PILs and frequent microflaring at sites of enhanced coronal heating in ARs was observed before by \cite{falc97} and \cite{moor99}. They proposed that the magnetic flux cancellation at these PILs was a process that could trigger/modulate the heating (rate of microflaring). However, this idea was not verified in want of high-cadence and high-resolution LOS magnetograms. Our high-resolution observations reveal that this process is, in fact, at work.

Our interpretation of the triggering of the fine-scale coronal flaring reported here is corroborated by a recent three-dimensional magnetohydrodynamic (MHD) simulation, although in a different magnetic setting from our case; subflares were triggered externally by reconnection of nearby emerging magnetic loops of the scale of the loops in our events \citep{arch14}. In the future, similar sophisticated MHD simulations might reproduce the external triggering of subflares in an AR corona, such as those observed here.  

In summary, we report new high-resolution observations of an external trigger mechanism of subflares in the braided magnetic structure of the AR corona observed by {\it Hi-C}. 
How common this external trigger mechanism for coronal microflares and subflares is in other ARs, and how important it is for coronal heating in general, remain to be explored by future observations by instruments having the resolution of {\it Hi-C} or better.


\acknowledgments
We thank the referee for constructive comments. The AIA and HMI data are courtesy of NASA/SDO and the AIA and HMI science teams. MSFC/NASA led the Hi-C mission and partners include the SAO in Cambridge, MA; LMSAL in Palo Alto, CA; the UCLan in Lancashire, England; and the LPIRAS in Moscow. S.K.T. and C.E.A. are supported by appointments to the NASA Postdoctoral Program at the NASA/MSFC, administered by ORAU through a contract with NASA. A.R.W. and R.L.M. are supported by funding from the LWS TRT Program of the Heliophysics Division of NASA's SMD.\\

\bibliographystyle{apj}
\bibliography{stiwari}

\end{document}